\documentclass[10pt,letterpaper]{article}

\usepackage{cogsci}
\usepackage{pslatex}
\usepackage{apacite}
\usepackage{mathtools}
\usepackage{tikz-cd}
\usepackage{fancyhdr} 

\fancyheadoffset{0cm}

\fancypagestyle{plain}{%
   \fancyhf{}%
   \fancyhead[R]{\thepage}%
}

\title{Interactions of Linguistic and Domain Overhypotheses in Category Learning}
 
\author{{\large \bf Luann C. Jung (luju@mit.edu)} \\
  Department of Computer Science and Engineering, MIT \\
  \\
  {\large \bf Haiyan Wang (hwang@ksu.edu)} \\
  Department of Statistics, Kansas State University}

\begin{document}

\maketitle

\begin{abstract}
For humans learning to categorize and distinguish parts of the world, the set of assumptions (overhypotheses) they hold about potential category structures is directly related to their learning process. In this work we examine the effects of two overhypotheses for category learning: 1) the bias introduced by the presence of linguistic labels for objects; 2) the conceptual `domain' biases inherent in the learner about which features are most indicative of category structure. These two biases work in tandem to impose priors on the learning process; and we model and detail their interaction and effects. This paper entails an adaptation and expansion of prior experimental work \cite{ivanovahofer} that addressed label bias effects but did not fully explore conceptual domain biases. Our results highlight the importance of both the domain and label biases in facilitating or hindering category learning.

\end{abstract}

\section{Introduction}

How do conceptual knowledge and language work together to affect the way we partition the world into categories? Some researchers \cite{lupyan2007} claim that verbal labels facilitate category learning by helping learners pick out relevant category dimensions. However, others \cite{brojde2011} have shown contrasting results that indicate that the presence of words can actually slow down learning. 

Recent work (Ivanova \& Hofer, 2020) developed a Bayesian computational model that reconciles these conflicting results. The authors argued that word labels impose a set of priors on the learning process. Thus, if the word-induced prior aligns with the true category structure, then words will facilitate learning. If the word-induced prior does not align with the true structure, they will make learning harder by biasing the learner toward irrelevant stimulus features. However, although this `label effect' is explored in depth by the authors, domain priors are not.

In this work, we alter and expand upon an existing Bayesian framework model \cite{ivanovahofer} for examining the effects of linguistic overhypotheses on category learning. Particularly, we examine the interaction between linguistic ``label-based'' priors and ``domain-based'' conceptual priors. By modeling the interactions between general concept priors and word-induced priors, we attain a more accurate representation of the priors involved in category learning.  For example, how do priors that drive the belief that shape is more likely to be a determinant of category than texture compare and interact with priors induced by labels? 

This work centers around examining and implementing these domain and label biases into a hierarchical Bayesian model. Through this model, we can better understand how learning is driven by interactions between language-independent and language-dependent biases. Specifically, we show that these two different biases both affect category learning, with the degree of each one's effect being variable and learnable.

\section{Prior Work}

Efforts have been made in prior works to describe the effect of labels on category learning. 

A study conducted by \citeA{lupyan2007} presented participants with two `alien' stimuli and had them learn to distinguish between which aliens to approach and which to avoid in a supervised learning setup. The distinguishing feature between the categories was shape, and the authors performed experiments by providing participants with labels for the aliens (`leebish' and `grecious') that aligned with category membership. These labels did not provide additional information about category membership to the participants (and thus could not be considered as features themselves). The study yielded results that the labels facilitated category learning and posited that the reason was because the labels helped to emphasize features where object variance aligned with category membership.

However, another study by \cite{brojde2011} communicated doubts about the beneficial nature of labels in category learning. Specifically, the authors found that providing shape-based labels with data that is categorized based on other dimensions such as texture or hue actually hindered category learning. This provided a contrasting view to the \textit{label advantage effect} of linguistic labels on learning. Hence, \citeA{brojde2011} concluded that instead of emphasizing category-distinguishing differences across all features, labels serve to bring the participants' attention towards dimensions that have been `historically relevant' for classification. 

To reconcile these two views, \citeA{ivanovahofer} proposed that verbal labels serve as overhypotheses for the learner in category learning. They assert that learners have a set of biases imposed by, for example, words whose meanings are characteristic of certain categories. Thus, when labels are used in the learning process they can either facilitate or hinder learning depending on whether the priors induced by the overhypotheses align with the true category structure. However, their model does not account for the potential conceptual domain biases that a learner may have. Specifically, in their model, the absence of label biases is considered a `no bias' situation, which they acknowledge is not truly representative of the learner due to the fact that even without any label biases the learner may still have inherent domain biases. 

\begin{figure}
    \centering
    \includegraphics[width=1\linewidth]{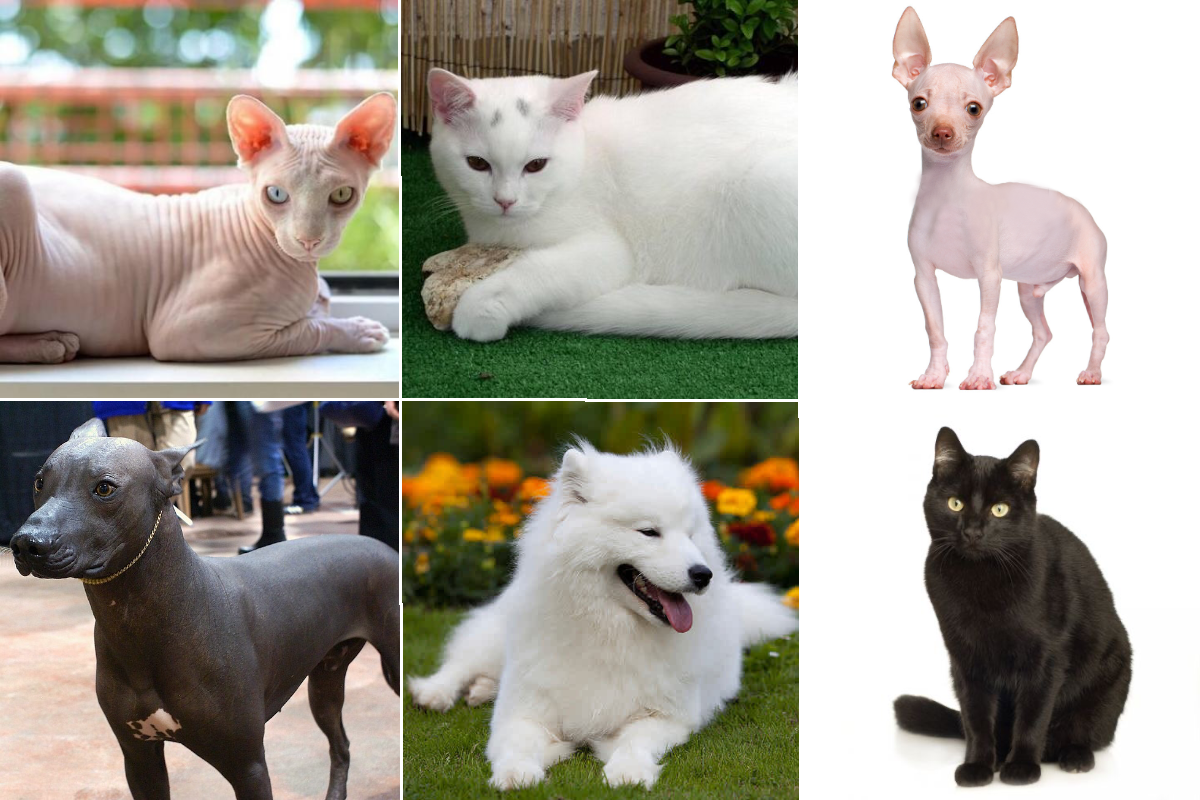}
    \caption{Six example objects for category learning that vary among numerous features (e.g. shape, texture, color, size, and position).}
    \label{fig:catdog}
\end{figure}

\section{Concept Overhypotheses}
The need for modeling of not only the label biases in category learning but also the domain biases is the impetus for this work. When a learner considers a category learning task, they tend to have some form of conceptual prior over which features are more indicative of category membership. 

An example of this can be demonstrated by looking at Fig \ref{fig:catdog}. Learners faced with these six objects for classification may have some inherent domain biases when learning the categories. For example, English speakers (a language where most nouns for objects are based on shape) tend to be more likely to classify objects by their shape than other features such as texture \cite{landau1988}.  In the example in Fig \ref{fig:catdog}, there are numerous features that the data can be partitioned by: shape, texture (hairless or furry), color (light or dark), size, and position (lying down, upright), among others. Focusing on each of these features as deterministic for categorization yields a different set of final categories. 

The conceptual domain biases that learners have when learning to categorize objects ultimately interplay with biases they develop when given linguistic labels for objects. One example of this is the case when a learner approaches the aforementioned category learning task with a conceptual bias that objects like these are more likely to be categorized by shape. Then, when provided with labels ``furry'' and ``hairless,'' they develop a label bias towards texture being the diagnostic feature (according to their understanding of the meanings of the labels). These biases (whether contrasting or constructive) both have effects on the learner's learning of category structure, especially in the early stages of supervised learning.

To better understand the processes by which humans perform these category learning tasks, we aim to account for the conceptual biases learners have about which dimensions are more informative of category membership in addition to any label-induced biases they may acquire. 

\section{Model Setup}

In this model, we consider a learning scenario where the goal is to learn to categorize object exemplars into $C$ disjoint categories. These exemplars possess $F$ perceptual dimensions (i.e. features) along which they may vary.  The categories themselves vary along at least one of the $F$ features. 


\begin{figure*}[!ht]
    \centering
\begin{tikzcd}
                      &                                       & \vec{\alpha_d} \arrow[d] \arrow[ld]     & \vec{\alpha_l} \arrow[d] \arrow[rd]     &                                       &                       \\
                      & domain \; bias_1 \arrow[d, "\omega"'] & domain \; bias_2 \arrow[rrd, "\omega"'] & label \; bias_1 \arrow[lld, "1-\omega"] & label \; bias_2 \arrow[d, "1-\omega"] &                       \\
                      & bias_1 \arrow[d]                      &                                         &                                         & bias_2 \arrow[d]                      &                       \\
                      & \Sigma_1 \arrow[ld] \arrow[rd]        &                                         &                                         & \Sigma_2 \arrow[ld] \arrow[rd]        &                       \\
{\mu_{1,1}} \arrow[d] &                                       & {\mu_{1,2}} \arrow[d]                   & {\mu_{2,1}} \arrow[d]                   &                                       & {\mu_{2,2}} \arrow[d] \\
{y_{1,1}}             &                                       & {y_{1,2}}                               & {y_{2,1}}                               &                                       & {y_{2,2}}            
\end{tikzcd}
    \caption{Illustration of the model structure with $F=C=2$. From the top down, $\omega$ and all $\vec{\alpha}$'s are fixed parameters; all biases, $\Sigma$'s, and $\mu$'s are stochastic variables; and all $y$ values are observed during learning.}
    \label{fig:modeltree}
\end{figure*}
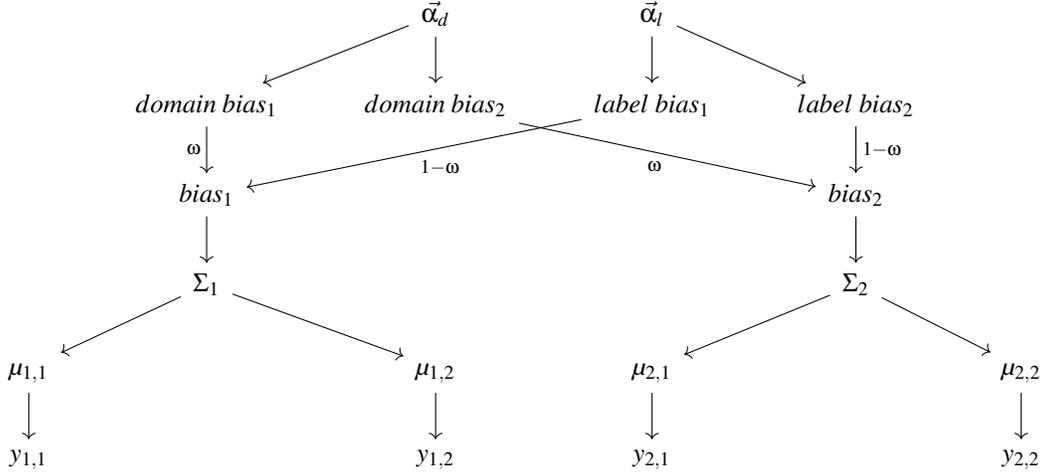

We focus on two types of bias that affect the way the learner evaluates and learns from the exemplars: 1) the bias induced in the learner by the presence of linguistic labels for the data; 2) the inherent domain bias of the learner. These domain and label biases were integrated into the model by considering a vector of bias parameters for each of the two biases. More specifically, the domain bias is parameterized by a vector $\vec{\alpha_d}$ that determines the learner's relative domain biases over each of the $F$ features. Likewise, the label bias is parameterized in the same manner by a vector $\vec{\alpha_l}$. 

The domain bias itself is a vector $\vec{p}$ containing $F$ values sampled from a Dirichlet distribution with parameter $\vec{\alpha_d}$. Similarly, the label bias is a vector $\vec{k}$ sampled from a Dirichlet distribution with aforementioned parameter $\vec{\alpha_l}$:
\begin{equation*}
\begin{split}
    \vec{p} = (p_1, ..., p_F)^T \sim \textrm{Dirichlet}(\vec{\alpha_d}), \\
    \vec{k} = (k_1, ..., k_F)^T \sim \textrm{Dirichlet}(\vec{\alpha_l}).
    \end{split}
\end{equation*}

These bias vectors $\vec{p}$ and $\vec{k}$ serve as overhypotheses during learning. The model also takes parameters $w$ and $s$ which serve to constrain how the domain and label biases interact. Particularly, $w$ and $s$ provide the mean and standard deviation that are used to sample a weight $\omega$ that will define a convolution of the domain and label biases;
\begin{equation*}
    \omega \sim \textrm{Normal}(w, s) \; \textrm{truncated at 0}.
\end{equation*}

By sampling $\omega$ from a truncated normal distribution centered at $w$ with lower bound zero, we are essentially choosing an estimate $w$ of how much belief the learner places on the domain bias and then adding a small amount of noise. This can capture to some extent the variability in how much more important a learner may consider the domain bias compared to the label bias or vice versa. 

Using the weight $\omega$, we can define $\omega \vec{p} + (1-\omega)\vec{k}$, a convolution of the two biases which is the total bias that the learner has when learning. Using $\omega$ and $1-\omega$ as the weights for the biases enforces that increased weight given to the domain bias corresponds to equally decreased weight for the label bias (and vice versa). The final bias values for each feature $i$ are transformed further using a power law function $r(p_i, k_i)$ scaled between -1 and 1 as follows:

\begin{equation*}
    r(p_i, k_i) = 2(\omega p_i + (1-\omega)k_i)^{\frac{1}{\gamma}} - 0.5).
\end{equation*}

Since the domain and label biases $\vec{p}$ and $\vec{k}$ were both drawn from Dirichlet distributions, they satisfy that $\sum_{i=1}^C{p_i}=1$ and $\sum_{i=1}^C{k_i}=1$. Accordingly, the convolution of $\vec{p}$ and $\vec{k}$ still maintains the same property $\sum_{i=1}^C{\omega p_i + (1-\omega)k_i}=1$. This allows for the domain and label biases to be considered with different weight, but still aligns with the so-called `conservation of belief' where if the learner is biased towards a particular feature then they are less likely to discover category distinctions based on other features. 

\begin{figure}[!th]
    \centering
    \begin{minipage}{0.5\linewidth}
    \centering
        \includegraphics[width=0.9\linewidth, height=0.17\textheight]{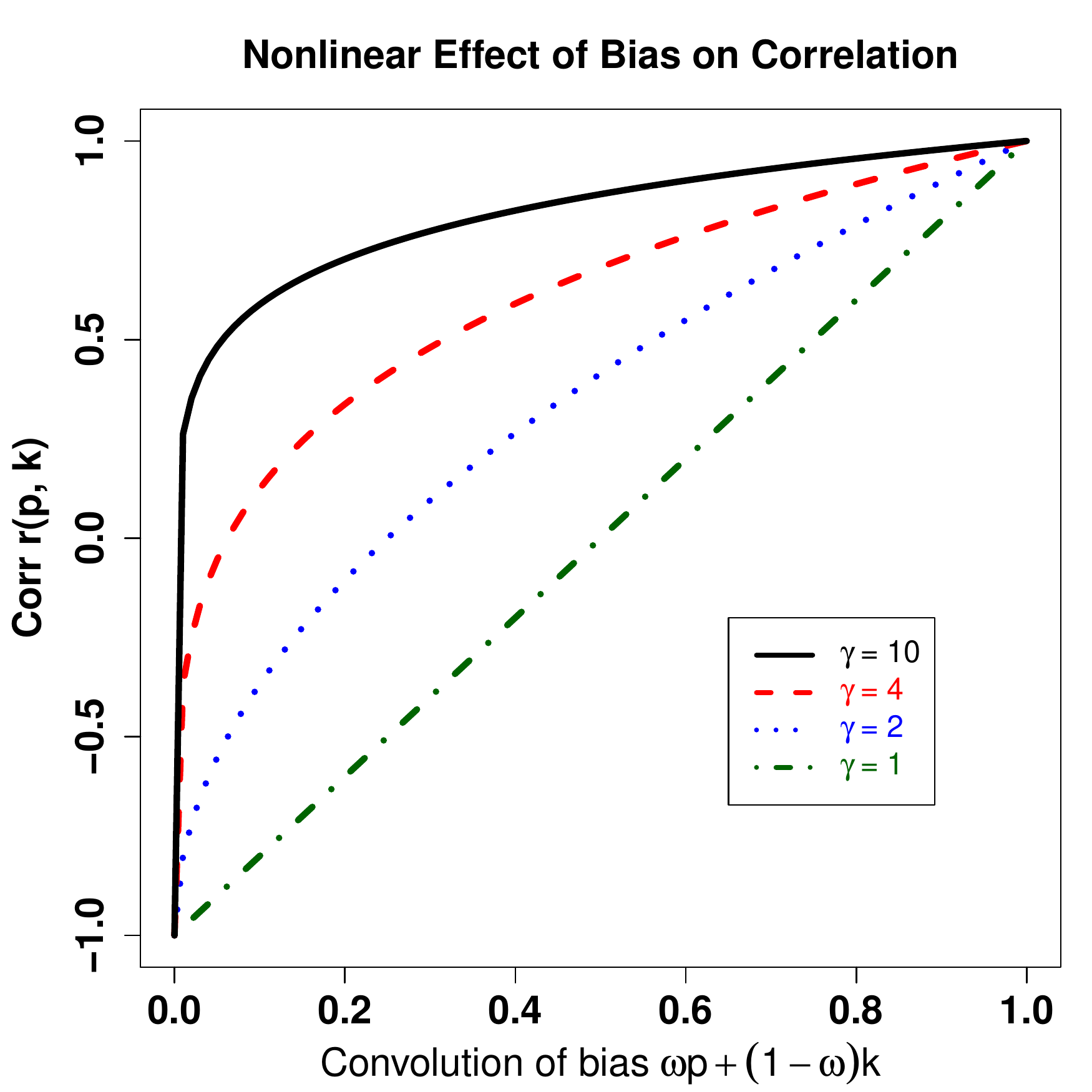}
    \end{minipage}
    \begin{minipage}{0.49\linewidth}
    \centering
        \includegraphics[width=.6\linewidth, height=0.17\textheight]{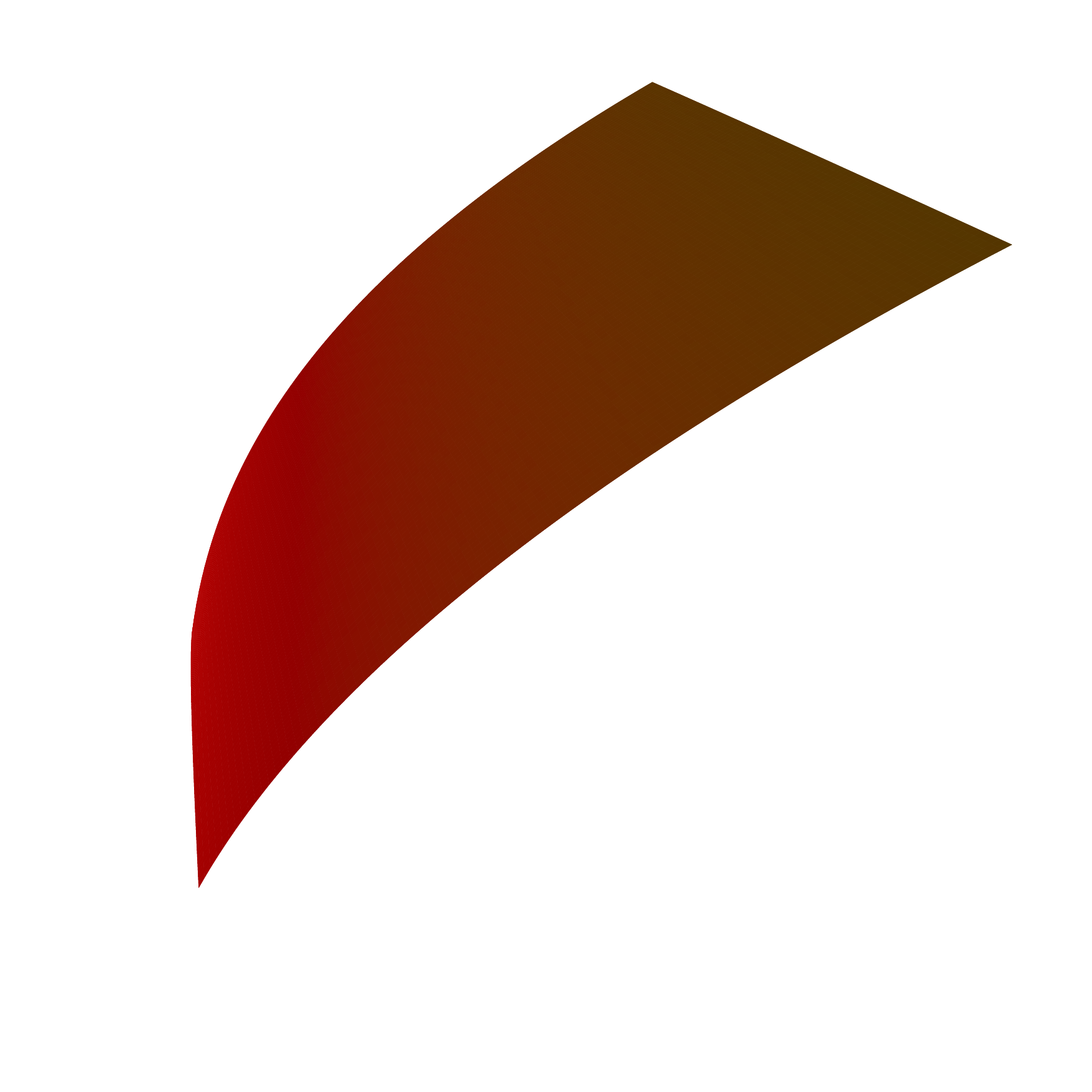}
    \end{minipage}
    \caption{\textbf{Left panel:} Nonlinear effect of different $\gamma$ values on the relationship between overall bias and correlations of category means. The rate of change for correlation values gets more extreme as bias approaches 0. Thus, a change in the bias closer to 0 results in a larger change in the distribution of the sampled category means. \; \textbf{Right panel:} A surface plot of the correlation between category means versus biases $p_1$ and $k_1$ for $p_1 \in [0.2, 0.5]$  and $k_1\in [0, 0.5]$. The resultant non-additive effect of the biases $p$ and $k$ is reflected in the nonlinear surface plot.}
    \label{fig:corrs}
\end{figure}

Using the function $r(p_i, k_i)$ to scale and transform the bias gives the correlation between the category means for the $i$th feature. More specifically, the correlation between the means of categories $m$ and $n$ for feature $i$ is defined as follows $\forall m, n \in 1, \ldots, C$:
\begin{equation*}
    \operatorname{corr}\left(\mu_{i, m}, \mu_{i, n}\right)=\left\{\begin{array}{ll}
r\left(p_i, k_{i}\right) & \text { if } m \neq n \\
1 & \text { otherwise. }
\end{array}\right.
\end{equation*}

\begin{figure*}[!t]
    \centering
    \begin{minipage}{.53\textwidth}
        \includegraphics[width=1\linewidth]{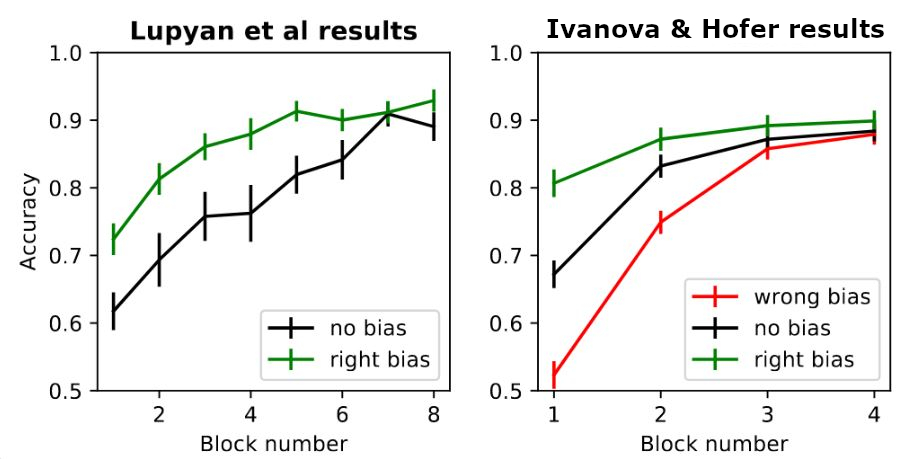}
    \end{minipage}
    \begin{minipage}{.46\textwidth}
        \includegraphics[width=1.05\linewidth]{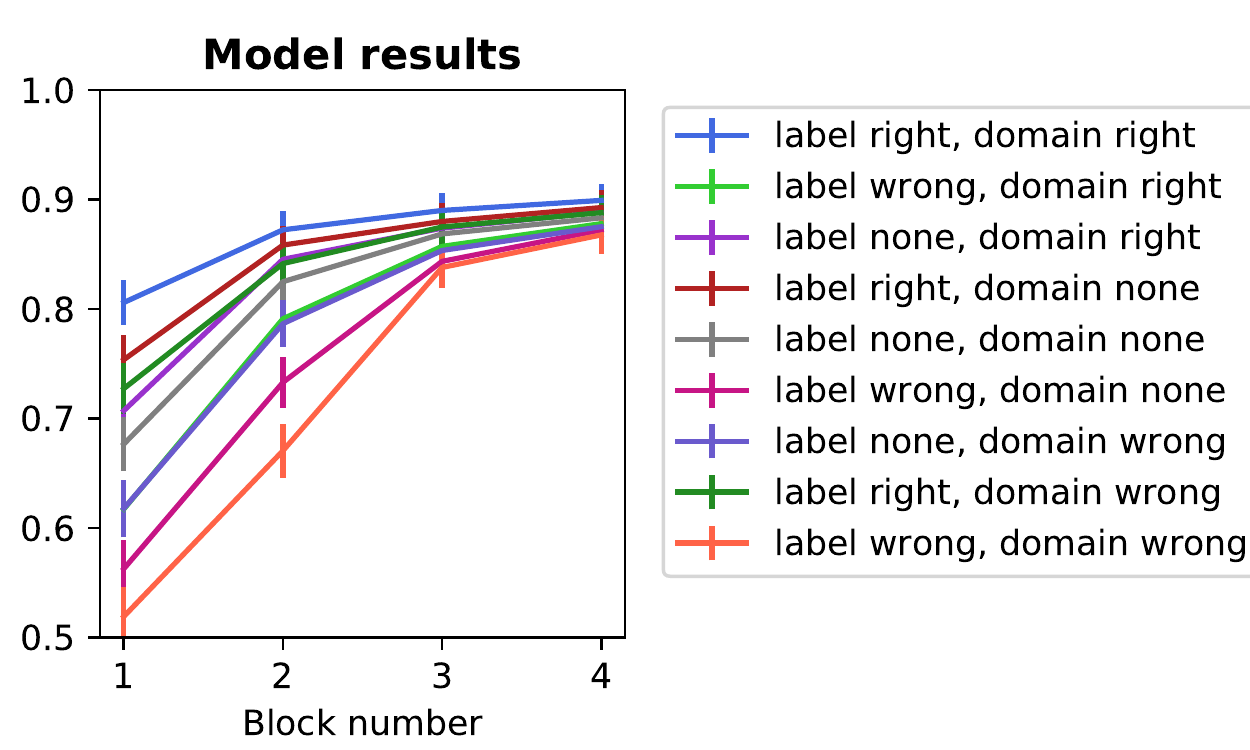}
    \end{minipage}
    \caption{Average accuracy of the model over 4 learning blocks compared to previous studies' results. The model  ($w=0.3, s=0.03$) results corroborate previous findings. Specifically, the presence of a `right bias' facilitates category learning (higher accuracies) while the presence of a `wrong bias' hinders learning (lower accuracies) compared to the `no bias' case. Additionally, the model shows that in general, a wrong label bias results in greater hindrance to learning compared to a wrong domain bias (the same applies for right biases). This occurs because for these settings of $w$ and $s$, the domain weight $\omega$ is less than the label weight $1-\omega$.} 
    \label{fig:modelingaccuracy}
\end{figure*}

The transformation $r$ serves to account for the need for nonlinear correlation values as discussed in \citeA{ivanovahofer}, where linear correlation values produce very little difference between right-bias, none-bias, and wrong-bias conditions. By using a power law function with parameter $\gamma$ we can specify how conservatively the learner generates hypotheses. When $\gamma$ is close to to 1 (linear), the learner is more likely to learn category structures that do not align with their biases. In contrast, larger values of $\gamma$ specify that the learner will primarily consider category structure hypotheses that align with their biases. The nonlinear transformation also augments the additive nature of the label and domain biases to result in a non-additive final effect (Fig \ref{fig:corrs}).

From  here, the model infers the category means and variances for each feature. Let $\vec{\sigma_i}$ be the vector of the variances of the $i$th feature for all $C$ categories:
\begin{equation*}
    \vec{\sigma_{i}}^2 = (\sigma_{i, 1}^2, ..., \sigma_{i, C}^2)^T.
\end{equation*}

Using the correlation values obtained from the biases, we can define a covariance matrix $\Sigma_i$ for each feature $i$:
\begin{equation*}
    \Sigma_i = diag(\vec{\sigma_{i}}) \cdot R_i \cdot diag(\vec{\sigma_{i}}),
\end{equation*}
where $R_i$ is the correlation matrix with entry $(m, n)$ being $\textrm{corr}(\mu_{i,m}, \mu_{i, n})$.

Then, we assume that the learner considers the category means for each feature as adhering to a Normal prior distribution with covariance matrix $\Sigma_i$:

\begin{equation*}
    \vec{\mu_i} = (\mu_{i,1}, ..., \mu_{i,C})^T \sim \textrm{Normal}(0, \Sigma_i)
\end{equation*}

These category means $\vec{\mu_i}$ and category variances $\vec{\sigma_{i}}^2$ (which are subject to some perceptual noise $\sigma_s^2$) are inferred by the learner based on observed category exemplars $y$. In this model, we make the simplifying assumption that the learner considers exemplars to be sampled from a Multivariate Normal distribution as follows:
\begin{equation*}
    \vec{y_i} = (y_{i,1},...,y_{i,C})^T \sim \textrm{MVN}(\vec{\mu_i}, diag(\vec{\sigma_{i}}^2) + \sigma_s^2I_C),
\end{equation*}
where $\vec{y_i}$ contains the learner's estimated values of feature $i$ in all categories.

Through this model setup, graphically visualized in Fig \ref{fig:modeltree}, we can explore the combined effect that conceptual domain priors and linguistic label priors have on category learning. 




\begin{figure*}
    \centering
    \setlength{\tabcolsep}{-10pt}
    \begin{tabular}{ccc}
        $w=0.2, s=0$ & $w=0.3, s=0.03$ & $w=0.5, s=0$\\
        \includegraphics[width=0.36\textwidth, height=0.23\textheight]{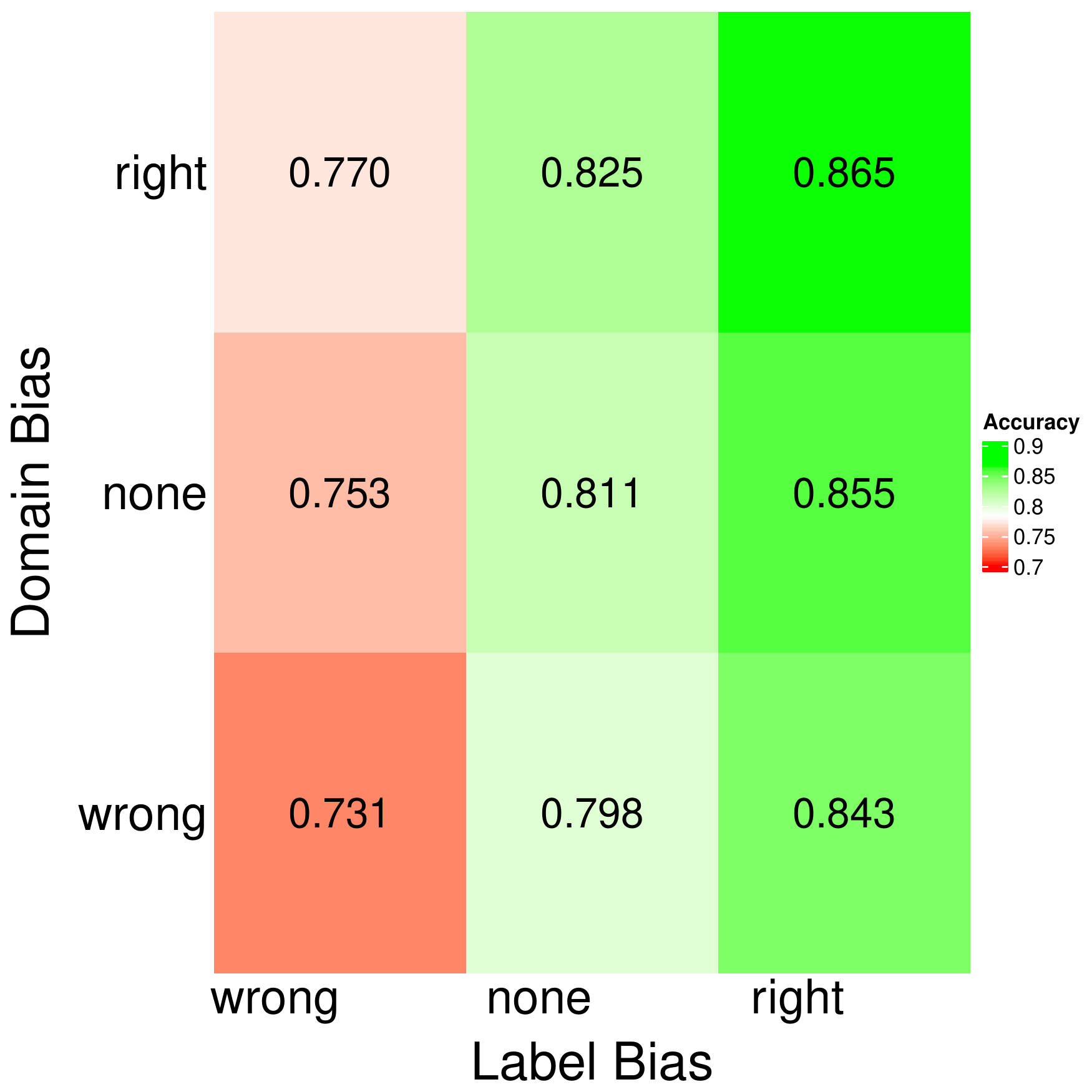} & \includegraphics[width=0.36\textwidth, height=0.23\textheight]{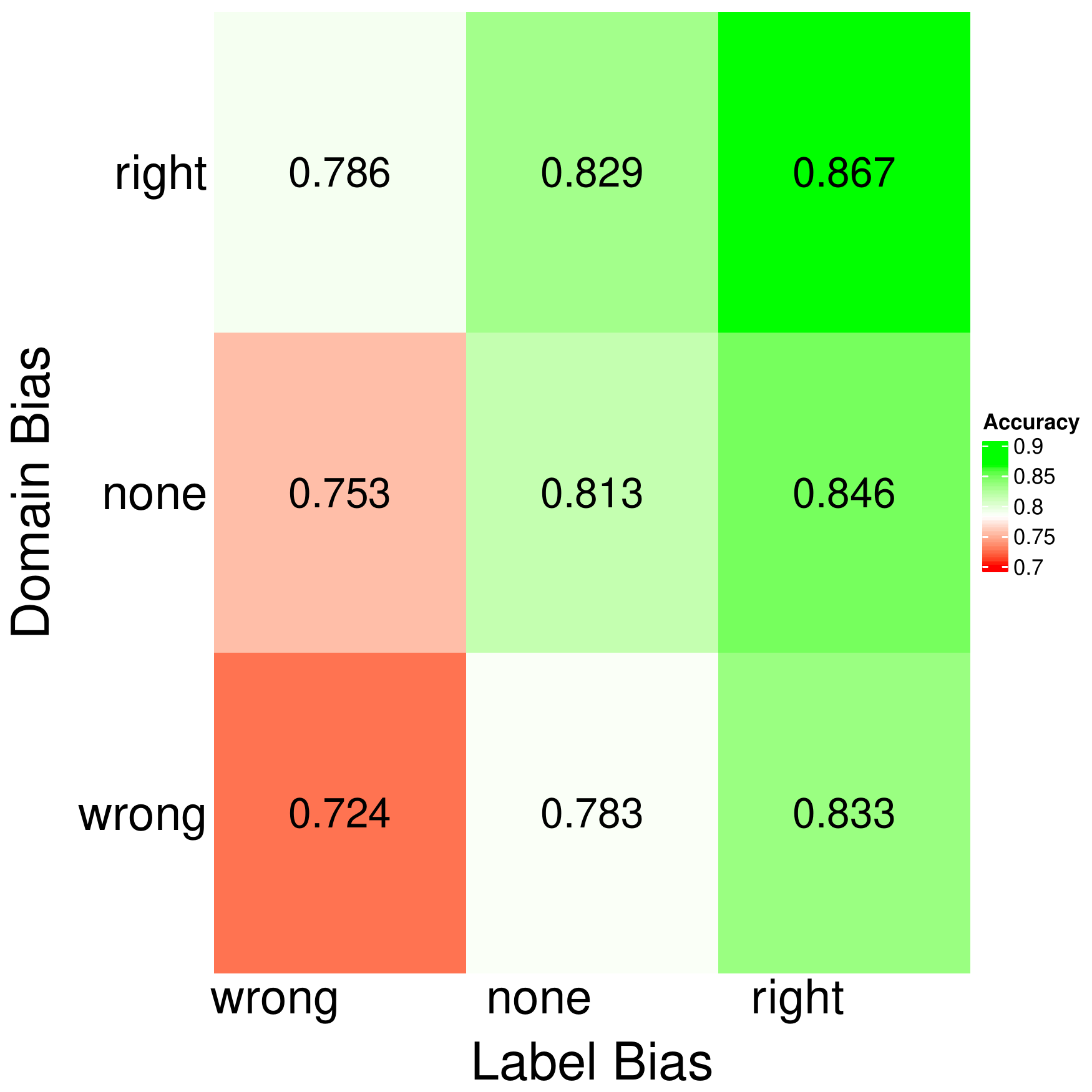} & \includegraphics[width=0.36\textwidth, height=0.23\textheight]{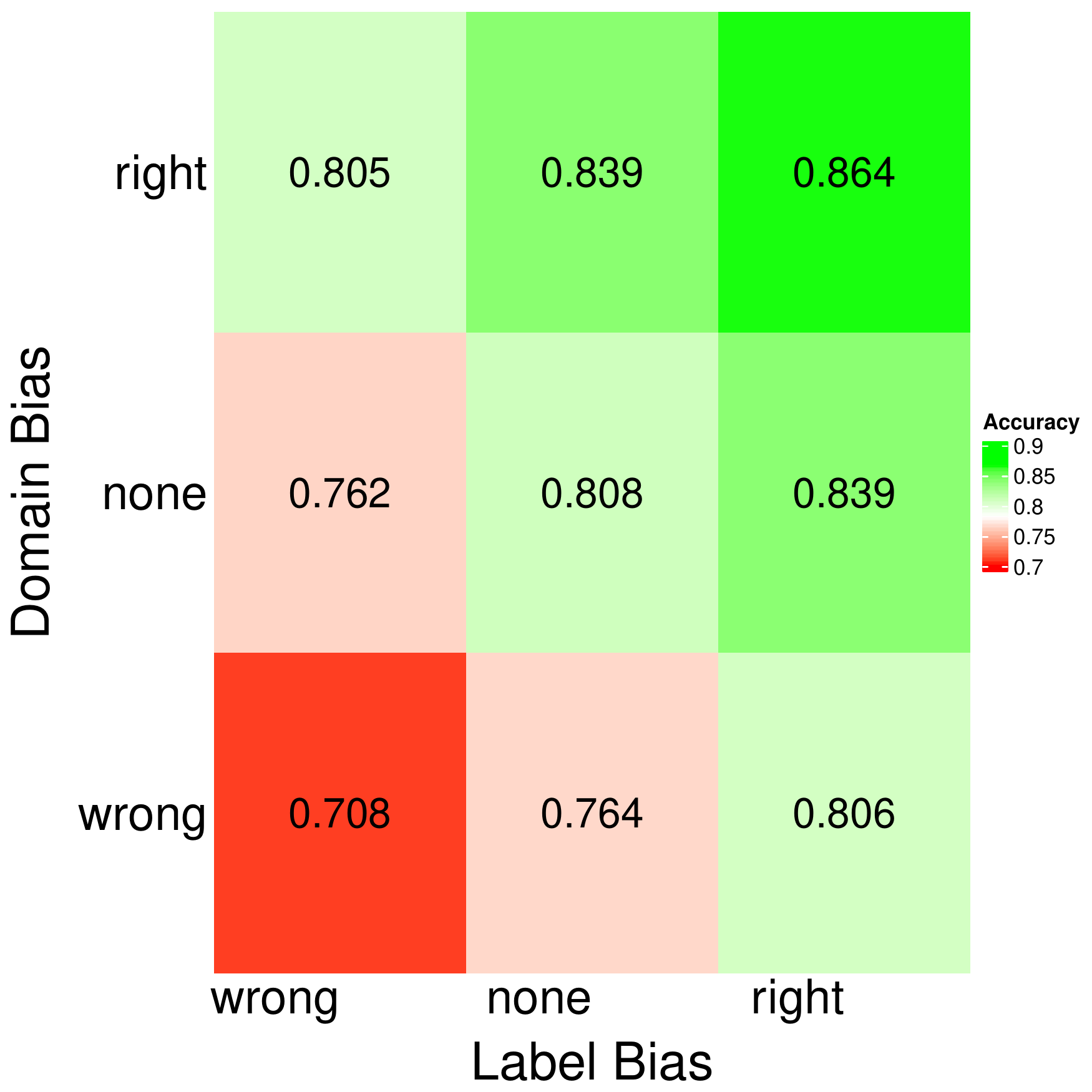}
    \end{tabular}
    \caption{Heatmaps showing the  average accuracy for each model for three different settings of $w$ and $s$. Note that when $w=0.5$ with no additional variance the domain and label biases are considered equally by the learner, so the accuracies are symmetric along the anti-diagonal. When $w=0.2$ or $w=0.3$, the learner places less belief in the domain bias compared to the label bias and this imbalance is reflected in the resulting accuracies.}
    \label{fig:heatmaps}
\end{figure*}

\section{Model Fitting}
\subsubsection{Data}
To fit the model, simulated data was generated in a manner to approximate the experiments in \citeA{brojde2011}. A total of 16 exemplars were generated with 8 in each of two categories. These simulated data varied along 2 dimensions, which, for simplification will subsequently be referred to as the ``texture'' and ``shape'' features. In the data, the shape feature was diagnostic of category. 

\subsubsection{Biases} The model was fitted 9 separate times to consider a variety of possible interactions between domain and label biases. We establish three levels of bias: the bias-induced prior aligns with the true category structure (``right bias''), the bias-induced prior does not align with the true category structure (``wrong bias''), and no priors are induced by the bias (``none bias''). The 9 fitted models consisted of all combinations of \{\texttt{right, none, wrong}\} domain bias and \{\texttt{right, none, wrong}\} label bias. 

\subsubsection{Markov Chain Monte Carlo} As in \citeA{lupyan2007}, \citeA{brojde2011}, and \citeA{ivanovahofer}, each model was given all 16 exemplars as observed data in every learning block. The models were fitted to the data using PyMC3, a Python-based probabilistic programming language \cite{salvatier2016}. 

In the fitting process, a Markov chain Monte Carlo (MCMC) simulation-based approach is used to obtain a Markov chain of values of the posterior distribution of the category means given the observed data. During the MCMC sampling, the No-U-Turn Sampler (NUTS) is used to generate posterior samples of the category means. NUTS is especially useful in our model due to its ability to handle many continuous parameters, a situation where other MCMC algorithms work very slowly. It computes the model gradients via automatic differentiation of the log-posterior density to find the regions where higher probabilities are present.

\section{Results}

We evaluated the performance of the 9 fitted models by classifying the observed exemplars into the two categories. This was done by comparing the exemplars' feature values to the learner's estimated feature means and standard deviations of each category. 

\begin{table}[!h]
    \centering
    \begin{tabular}{ll}
        \hline
         Bias & $\vec{\alpha}$ \\
         \hline
         Right & (1, 10) \\
         None & (10, 10) \\
         Wrong & (10, 1)
    \end{tabular}
    \caption{Settings of $\vec{\alpha}$ for each class of bias}
    \label{tab:biastypes}
\end{table}

In the model, some parameters were held as constants. These were the perceptual noise $(\sigma_s^2=1)$ and the nonlinear transform parameter $(\gamma=10)$.

We also established settings of $\vec{\alpha_d}, \vec{\alpha_l} \in$ \{\texttt{(1, 10), (10, 1), (10, 10)}\} according to Table \ref{tab:biastypes} to represent the domain and label biases. In setting these $\vec{\alpha}$s we are choosing a fixed representation for the overhypotheses. However, it is notable that in principle, these overhypotheses can be learned from data \cite{kemp2007}.

\subsubsection{Comparison of Models}

The model was able to demonstrate the effects of the domain and label biases on learning. 

At a basic level, holding the domain bias constant at `none' produced results that align with those in \citeA{ivanovahofer}, where we see that `right label bias` facilitated learning while `wrong label bias` hindered it (Fig \ref{fig:modelingaccuracy}). 

In addition, the other models fitted with various label and domain biases for $w=0.3, s=0.03$ (Fig \ref{fig:modelingaccuracy}) show that the label bias has a stronger effect (whether positive or negative) on learning. This aligns with the expected results due to the tested settings of $w$ and $s$ generating a domain weight $\omega$ that is less than the label weight $1-\omega$.

When looking at the average accuracies yielded by each of the 9 fitted models, we see a reflection of the hypothesized effects of the domain and label biases (Fig \ref{fig:heatmaps}). When the two biases are considered with equal weight by the learner, the resulting accuracy matrix for the different combinations of domain and label bias is symmetric along the anti-diagonal (Fig \ref{fig:heatmaps}). This aligns with the expectation that when the learner considers the two biases equally while learning, changes in one bias have very similar effects to equivalent changes in the other bias on overall accuracy. 

\begin{figure*}
    \centering
    \setlength{\tabcolsep}{-10pt}
    \begin{tabular}{ccc}
        $w=0.2, s=0$ & $w=0.3, s=0.03$ & $w=0.5, s=0$\\
        \includegraphics[width=0.36\textwidth]{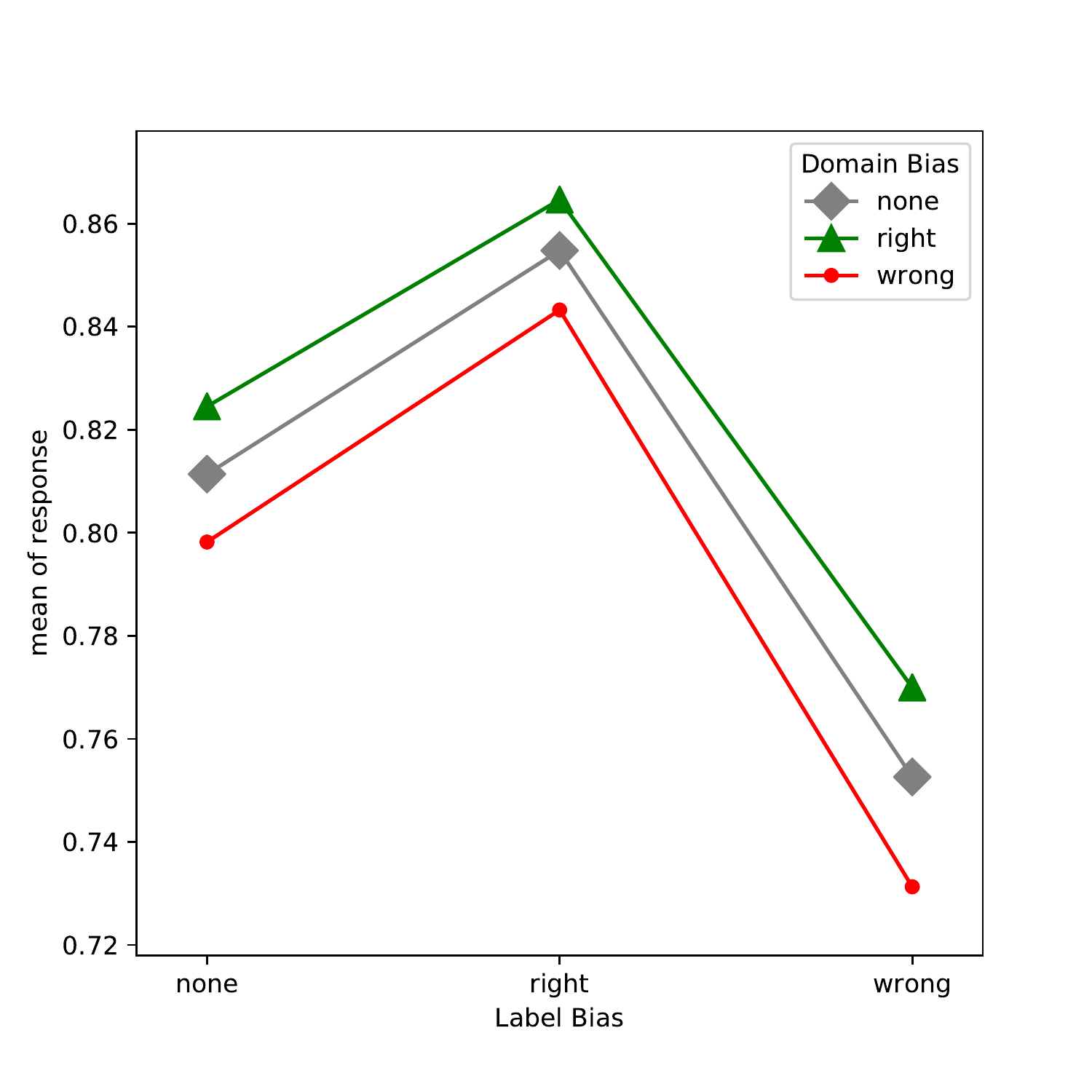} & \includegraphics[width=0.36\textwidth]{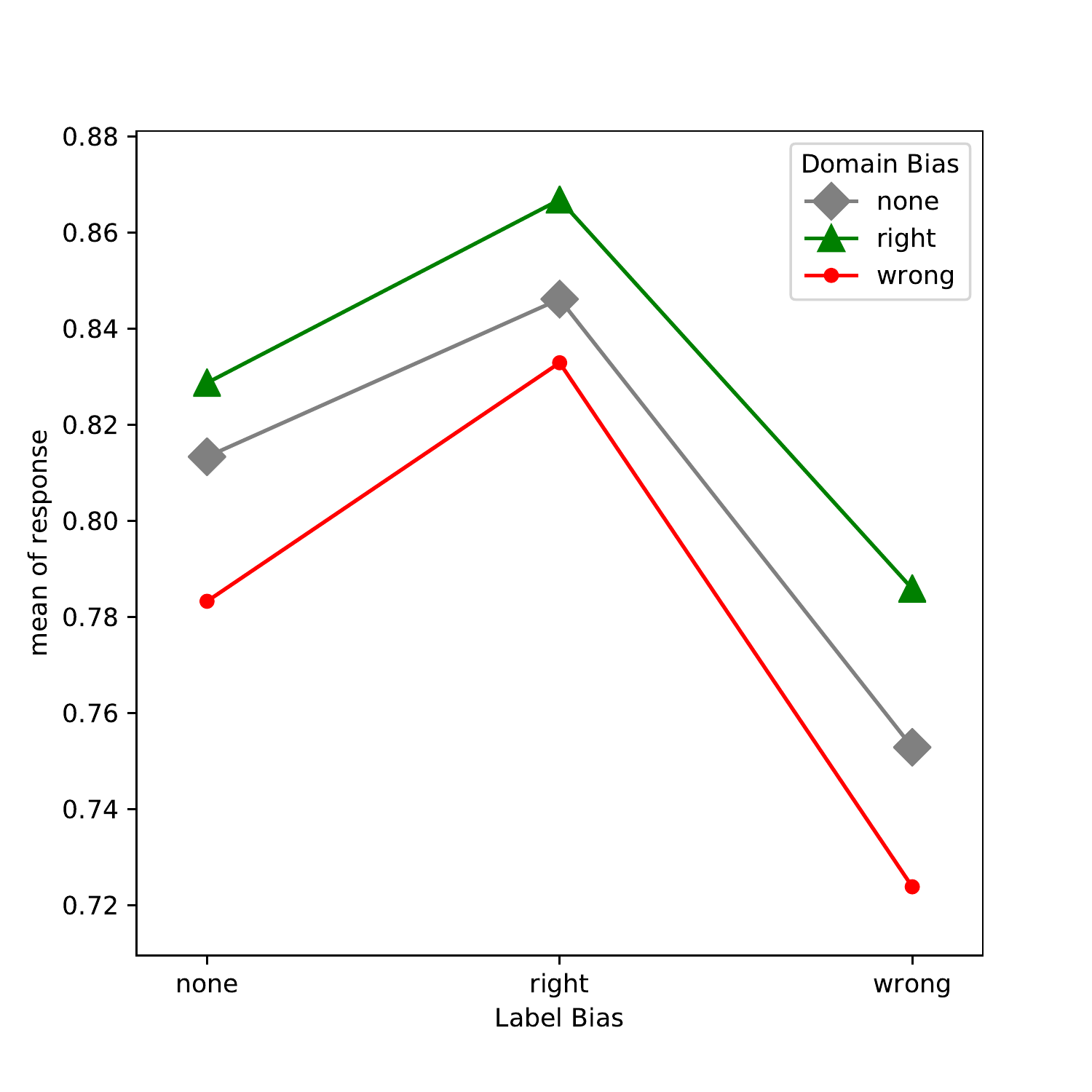} & \includegraphics[width=0.36\textwidth]{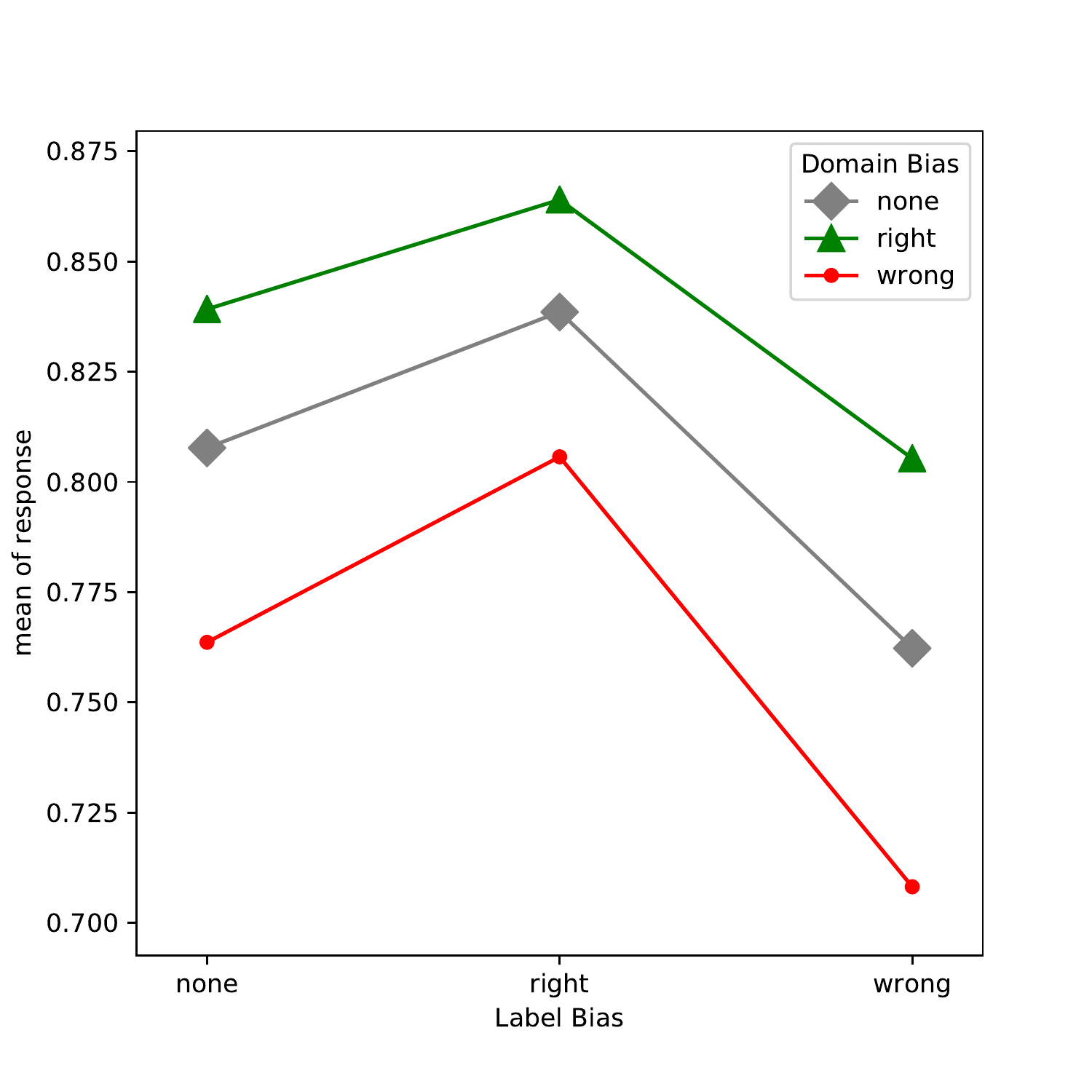}
    \end{tabular}
    \caption{Interaction plots for three different settings of $w$ and $s$ showing average accuracy by model. When $s=0$, the lines are basically parallel and indicate minimal interaction between the domain and label biases. However, of particular interest is the introduction of a nonzero $s$ value (as is the case when $w=0.3$). In this case, the additional variability in $\omega$ makes the lines in the plot begin to deviate from parallel.}
    \label{fig:interactions}
\end{figure*}

However, for unequal weights (e.g. when $w=0.2$ and $w=0.3$ in Fig \ref{fig:heatmaps}), the effect of changes in one bias (label bias in the example) is shown to be stronger than equivalent changes in the other bias. Additionally, the inclusion of a nonzero $s$ when $w=0.3$ introduces noise to the bias weights -- a scenario more representative of a real human learner.

\begin{figure}
    \centering
     \includegraphics[width=1\linewidth]{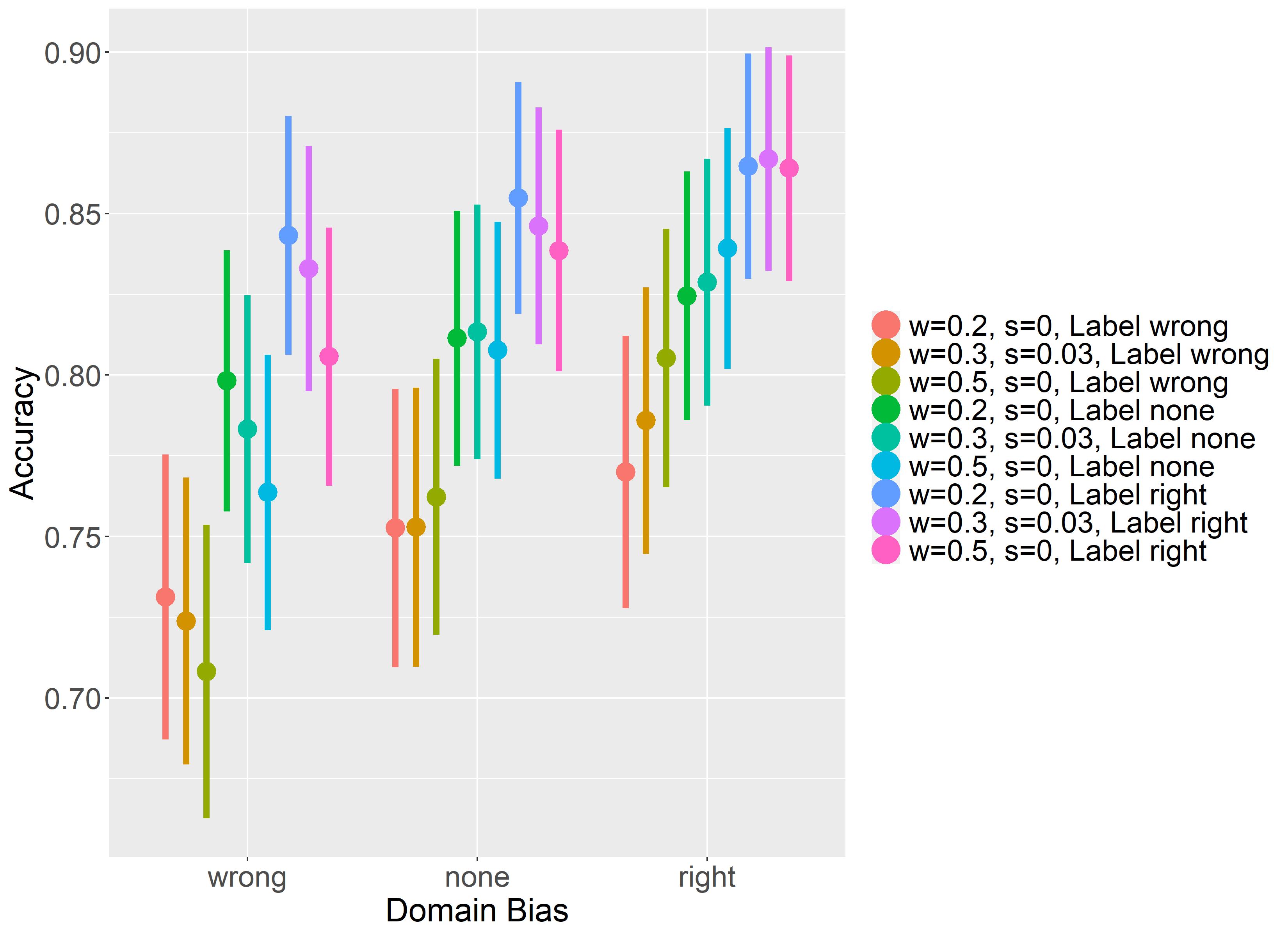}
        \caption{Dot plot of the average model accuracies and standard errors for each of the 9 model fittings for three different settings of ($w, s$): (0.2, 0), (0.3, 0.03), and (0.5, 0). We can see that the different values of $w$ have slightly different effects on the magnitude and direction of the accuracy trends for any fixed domain bias and label bias.} 
        \label{fig:dotplot}
\end{figure}

Finally, another important observation from the results is demonstrated in Fig \ref{fig:dotplot}. Within each particular bias type and class (e.g. right domain bias), changing the other bias from `wrong' to `none' to `right' generally increases accuracy. However, within a particular setting of both bias types and classes (e.g. wrong domain bias and wrong label bias), the accuracy trends vary when parameter $w$ is changed. For example, when domain bias and label bias are both wrong, increasing $w$ (and therefore $\omega$, the degree to which the learner believes their domain bias) decreases accuracy. However, when there is no domain bias and wrong label bias, increasing $w$ has negligible effect on the accuracy. These results all align with our initial hypotheses about the behavior of the domain and label biases.

The model also attempts to capture the interactions between the domain and label biases. The convolution of the two biases and the decision to put both on the same level of the Bayesian hierarchical framework was informed by adhesion to the principle of `conservation of belief' where increased belief or weight placed in one type of bias should result in a corresponding decrease in others. Although the initial convolution was additive, the final contribution of the combined biases to the correlations between the category means is not linear (Fig \ref{fig:corrs}). This could potentially lead to interaction effects between the two biases. However, such interactions were minimal in our simulation results.

\begin{table*}[t]
\centering
\setlength{\tabcolsep}{2.5pt}
\renewcommand{\arraystretch}{1}
\begin{tabular}{l|r|cc|cc|cc}
  \hline
 & &\multicolumn{2}{|c|}{$w=0.2, s=0$}&\multicolumn{2}{c|}{$w=0.3,
s=0.03$}& \multicolumn{2}{c}{$w=0.5, s=0$} \\
  &    & GEE      & GLMER   & GEE     & GLMER   & GEE     & GLMER \\
Effect  & Df &  p-value & F-value & p-value & F-value & p-value & F-value \\
  \hline
Block &   1 & 0 & 1896.8 & 0 & 1824.6 & 0 & 1784.3 \\
  Label Bias &   2 & 0 & 249.2 & 4.00e-15 & 205.2 & 1.83e-09
& 124.8 \\
  Domain Bias &   2 & 0.016 & 22.4 & 5.49e-05 & 57.9 & 3.33e-10 & 121.5 \\
  block:Label Bias &   2 & 5.13e-05 & 65.9 & 3.69e-04 & 55.2 &
0.099 & 16.5 \\
  block:Domain Bias &   2 & 0.941 & 0.5 & 0.177 & 12.0 &
0.021 & 25.0 \\
  Label Bias:Domain Bias &   4 & 0.931 & 1.3 & 0.741 & 2.7 &
0.583 & 3.8 \\
   \hline
\end{tabular}
\caption{Results from mixed effects logistic regression and
generalized estimation equation with logit link and exchangeable
correlation structure. The interaction effect of label and domain bias
is not significant from either the GEE and GLMER models.}
\label{tab:testresult}
\end{table*}

As seen in Fig \ref{fig:interactions}, changes in the model parameters $w$ and $s$ (and therefore weight parameter $\omega$) demonstrate mildly different levels of interaction between domain and label biases. For the cases when $s=0$ ($w=0.2$ and $w=0.5$), the lines in the plot do not intersect at any point. This indicates that there are no significant interactions between the domain and label biases when the learner's belief in their domain bias is not subject to any variation. 

Finally, parallel to the procedure used in \citeNP{ivanovahofer}, we simulated individual-level predictions with 75 participants per setting of domain and label biases class combinations in each learning block. These were analyzed using a mixed effects logistic regression model that was fitted using the \texttt{glmer} command from the \texttt{lme4} package \cite{bates2015} with formula:
\begin{equation*}
    accuracy \sim (label + domain + block)^2 + (1 | object),
\end{equation*}
and a generalized estimating equation (GEE) model with observations from the same object having an exchangeable correlation structure and formula:
\begin{equation}
    accuracy \sim (block+label + domain)^2
\end{equation}

These two models have identical correlation matrices, but GEE provides a test for the fixed effects while \texttt{glmer} does not. We used both models to assess whether the interaction effect between the label and domain biases is significant. 


We concluded that the interaction effect for the model fitted with different settings of $(w,s)$ was not statistically significant based on both the $p$-values from the GEE model and the F-statistics from the mixed model as shown in Table \ref{tab:testresult}. This quantitatively confirms the patterns we observed in the interaction plot (Fig \ref{fig:interactions}) where there are not significant interactions between domain and label biases (at least at the level of weight variation $s$ tested). Although the examples modeled did not show significant interaction between biases, the use of more complex data and therefore more complex bias priors could very likely amplify any domain-label interactions. We did observe significant effects for block \& label, block \& domain, and main effects of block, label, and domain. Some of these effects were also reported by \citeNP{brojde2011}.

\section{Discussion}

\subsubsection{Domain weight flexibility} In this work, the additional parameters $w$ and $s$ that further parameterize the domain bias weight $\omega$ served an important role in making the learner's consideration of domain vs label biases flexible. It appears very reasonable that these parameters are also learnable in category learning tasks -- i.e. learners learn to either emphasize or discount their domain bias as they go through the supervised learning process, thus altering their $\omega$.

\subsubsection{Importance of parameters $w$ and $s$} 
Different learners have different inherent degrees of belief in the two biases. This variability is able to be partially captured by the noise introduced by the $s$ parameter. Hence, the capability to adjust $s$ is important for estimating different types of learners.

Beyond these smaller variations between learners, learners may also vary wildly in their preference for domain or label belief. One key example of this is in comparing adult learners to infant learners. Infants will have less developed understanding of word meanings and therefore the label bias effect may be negligible for them (high $\omega$). On the contrary, adults may be strongly influenced by label biases because they understand the structure underlined by the labels (high $1-\omega$). The model can account for these differences by adjusting the magnitude of $w$, which is where the distribution of $\omega$ is centered, to model learners with vastly different beliefs in the two biases.

\subsubsection{Limitations} It is important to acknowledge the limitations of this model. Although the model detailed in this work provides an implementation of the domain and label biases, additional data collection to tune the model's assumptions was outside of the scope of this project and thus was not conducted.

Additionally, the model is just one implementation of the possible combined behaviors of the label and domain biases. Hence, alternative implementations may be worth investigating to compare and contrast results.

\subsection{Future Research}
Given the limitations detailed previously, a key future direction to consider for this work is adjusting the structure of the Bayesian hierarchical model. In this work, we discussed that the convolution of the domain and label biases is additive. Despite this additive implementation, we use a combination of the nonlinear transformation and the learners' variations $s$ in the degree to which they believe in their domain biases to introduce potential interactions between the biases. However, as we saw in Fig. \ref{fig:interactions}, these interactions are still minimal. One way to further model more complex interactions between the domain and label biases when category learning is to consider the domain bias on a separate level from the label bias. This is particularly relevant given our assumption about the label bias for this work: that the label biases are generated by the learner's understanding of word meanings and their alignment with certain dimensions. Since the label bias depends in part on the learner's understanding of the domain, the label biases and domain biases may be more interconnected than defined in this model. Considering the domain bias on a different level of the hierarchical framework or using it to put priors on the label bias (which would then propagate down to generate priors for the category mean correlations) would be interesting ways to implement the domain and label biases differently.

Finally, data collection to compare the simulated results of this model with human data is an important next step for further verifying the conclusions generated in this work. 

Overall, this paper clarifies and conceptualizes the importance of considering both label and domain biases in category learning. By representing the biases as overhypotheses that impose priors over the learning process, we are able to model the effects each bias has on learning, as well as show how they interact.

\section{Acknowledgments}

Thanks to Anna Ivanova and Matthias Hofer for participating in discussion at the early stages of this project.

\nocite{brojde2011}
\nocite{ivanovahofer}
\nocite{kemp2007}
\nocite{landau1988}
\nocite{lupyan2015}
\nocite{lupyan2007}
\nocite{salvatier2016}

\bibliographystyle{apacite}

\setlength{\bibleftmargin}{.125in}
\setlength{\bibindent}{-\bibleftmargin}

\bibliography{CogSci_Template}

\end{document}